# Evaluation of potential energy of interaction between molecules using one-range addition theorems for Slater type orbitals and Coulomb-Yukawa like correlated interaction potentials

I.I. Guseinov

*Department of Physics, Faculty of Arts and Sciences, Onsekiz Mart University, Çanakkale, Turkey*

**Abstract**

Using one-range addition theorems for noninteger n Slater type orbitals and Coulomb-Yukawa like correlated interaction potentials with noninteger indices obtained by the author with the help of complete orthonormal sets of $\psi^\alpha -$ exponential type orbitals $(\alpha = 1, 0, -1, -2, ...)$, the series of expansion formulas are established for the potential produced by molecule, and the potential energy of interaction between molecules through the radius vectors of nuclei of molecules, and the linear combination coefficients of molecular orbitals. The formulae obtained are useful especially for the study of interaction between atomic-molecular systems containing any number of closed and open shells when the Hartree-Fock-Roothaan and explicitly correlated methods are employed. The relationships obtained are valid for the arbitrary values of indices and screening constants of orbitals and correlated interaction potentials.

## 1. Introduction

In molecular quantum-mechanical calculations, the choice of reliable basis atomic orbitals is of prime importance since the quality of molecular properties depends on the nature of these orbitals. During the past few years, a pragmatic preference has developed for Gaussian type orbitals (GTO) basis functions. This is motivated by the practical requirement for easy and rapid evaluation of multicenter integrals [1]. Unfortunately, the GTO basis functions fail to satisfy two mathematical conditions for atomic electronic distributions, namely, the cusp condition at the origin [2] and exponential decay at long range [3]. The Slater type orbitals (STO) would be desirable for basis sets because they satisfy these conditions. Unfortunately, the STO functions are not orthogonal with respect to the principal quantum numbers that creates some difficulties in molecular electronic structure calculations. Thus, the neccessity for using the complete orthonormal sets of $\psi^\alpha -$ exponential type orbitals $(\psi^\alpha - ETO, \alpha = 1, 0, -1, -2, ...)$ as basis functions arises [4]. In previous papers [5] and [6], with the help of $\psi^\alpha - ETO$, the different sets of unsymmetrical and symmetrical one-range addition theorems have been established for the expansion of multicenter charge densities of STO and Coulomb-Yukawa like correlated interaction potentials (CIP) with integer and noninteger indices, respectively. We notice that the use of CIP with noninteger indices considerebly improves the effectiveness of calculations (see. Ref. [7] and reference quoted therein)

The purpose of this work is to obtain the series of expansion formulae for the potential produced by molecule and the energy of interaction between molecules by the use of one-range addition theorems presented in Refs. [5] and [6] when the STO and the Coulomb-Yukawa like CIP approximations in Hartree-Fock-Roothaan (HFR) and explicitly correlated methods are employed. The results presented are especially useful for the investigation of interaction between atomic-molecular systems.

## 2. Definitions and basic formulas

The noninteger $n^*$ and integer $n$ ($for\, n^* = n$) STO (ISTO and NISTO) basis sets and the Coulomb-Yukawa like CIP with noninteger $\mu^*$ and integer $\mu$ ($for\, \mu^* = \mu$) indices (ICIP and NICIP) (see Ref. [7]) used in this work are defined as

$$\chi_{n^*lm}(\zeta,\vec{r}) = (2\zeta)^{n^*+1/2}[\Gamma(2n^*+1)]^{-1/2} r^{n^*-1} e^{-\zeta r} S_{lm}(\theta,\varphi) \tag{1}$$

$$h^{\mu^*\nu\sigma}(\xi,\vec{r}) = \left(\frac{4\pi}{2\nu+1}\right)^{\frac{1}{2}} r^{\mu^*-1} e^{-\xi r} S_{\nu\sigma}(\theta,\varphi). \tag{2}$$

Here, $\zeta$ and $\xi$ are the scaling paramaters of STO and potential, respectively ($\zeta > 0\, and\, \xi \geq 0$).

The spherical hormonics $S_{lm}(\theta,\varphi)$ are determined by relation

$$S_{lm}(\theta,\varphi) = P_{l|m|}(\cos\theta)\Phi_m(\varphi), \tag{3}$$

where $P_{l|m|}$ are the normalized associated Legendre functions and

for complex spherical harmonics $(S_{lm} \equiv Y_{lm})$

$$\Phi_m(\varphi) = \frac{1}{\sqrt{2\pi}} e^{im\varphi}, \tag{4}$$

for real spherical harmonics

$$\Phi_m(\varphi) = \frac{1}{\sqrt{\pi(1+\delta_{m0})}} \begin{cases} \cos|m|\varphi & for \quad m \geq 0 \\ Sin|m|\varphi & for \quad m < 0. \end{cases} \tag{5}$$

The NISTO are the orthogonal with respect to the quantum numbers $l$ and $m$:

$$\int \chi^*_{n^*lm}(\zeta,\vec{r}) \chi_{n'^*l'm'}(\zeta',\vec{r}) dv = \delta_{ll'}\delta_{mm'} \frac{\Gamma(n^*+n'^*+1)}{\left[\Gamma(2n^*+1)\Gamma(2n'^*+1)\right]^{1/2}} (1+t)^{n^*+1/2}(1-t)^{n'^*+1/2}, \tag{6}$$

where $t = (\zeta-\zeta')/(\zeta+\zeta')$.

We notice that the definition of phases in this work for the complex spherical harmonics $\left(Y_{lm}^* \equiv Y_{l-m}\right)$ differ from the Condon-Shortley phases [8] by the sign factor $(-1)^m$.

## 3. Potential of electric field produced by charges of molecule

The operator of potential produced by charges of molecule (see Fig. 1) is defined as

$$\hat{\varphi}(\vec{r}_{og}) = \sum_b Z_b \hat{h}^{\mu\nu\sigma}(\xi, \vec{r}_{bg}) - \sum_{i=1}^{N} \hat{h}^{\mu\nu\sigma}(\xi, \vec{r}_{gi}), \qquad (7)$$

where $\vec{r}_{og}$ and $\vec{r}_{bg}$ are the radius-vectors of the point $g$ with respect to the origin of the molecular coordinate system and to the nuclei of molecule, respectively ($b \equiv a, c, ...$), $Z_b$ is the charge of nucleus b, N is the number of electrons and $r_{gi}$ is the distance to the $i^{th}$ electron of molecule. The Coulomb-Yukawa like NICIP operator occurring in (7) is defined by Eq. (2).

With the calculation of average expectation value of operator (7) one can use the method set out in previous paper [9]. Then, we obtain for the potential produced by molecule which has the multideterminantal single electron configuration states with any number of closed and open shells the following relations:

$$\varphi(\vec{r}_{og}) = \sum_b Z_b h^{\mu^*\nu\sigma}(\xi, \vec{r}_{bg}) - 2\sum_{i=1}^{n} f_i h_i^{\mu^*\nu\sigma}(\xi, \vec{r}_{og}) \qquad (8)$$

$$h_i^{\mu^*\nu\sigma}(\xi, \vec{r}_{og}) = \int u_i^*(\vec{r}_1) \hat{h}^{\mu^*\nu\sigma}(\xi, \vec{r}_{g1}) u_i(\vec{r}_1) dv_1. \qquad (9)$$

Here $n = n_c + n_0$ is the number of occupied orbitals bilonging to closed ($n_c$) and open ($n_0$) shells, and $f_i$ is the fractional occupancy of shell i. The molecular orbitals $u_i$ occurring in Eq. (9) are defined as linear combination of NISTO determined by Eq. (1):

$$u_i = \sum_{p^*} \chi_{p^*} C_{p^*i}, \qquad (10)$$

where $p^* \equiv n^* l m$.

The linear combination coefficients $C_{p^*i}$ can be determined by solving the combined HFR equations for molecule (see Ref.[9]).

In order to evaluate the integral (9) in the MO LCAO approximation we take into account (10). Then, we obtain:

$$h_i^{q^*}(\xi, \vec{r}_{og}) = \sum_{p^*} \sum_{p'^*} C_{p^*i}^* C_{p'^*i} \Lambda_{p^*p'^*q^*}^{acg}(\zeta, \zeta', \xi) \qquad (11)$$

$$\Lambda_{p^*p'^*q^*}^{acg}(\zeta, \zeta', \xi) = \int \chi_{p^*}^*(\zeta, \vec{r}_{a1}) \chi_{p'^*}(\zeta', \vec{r}_{c1}) h^{q^*}(\xi, \vec{r}_{g1}) dv_1, \qquad (12a)$$

$$= \frac{1}{(2\nu+1)^{1/2}} \frac{[\Gamma(2\mu^*+1)]^{1/2}}{(2\xi)^{\mu^*+1/2}} I^{acg}_{p^* p'^* q^*}(\zeta, \zeta', \xi), \tag{12b}$$

where $p^* \equiv n^*lm$, $p'^* \equiv n'^*l'm'$, $q^* \equiv \mu^*\nu\sigma$ and

$$I^{acg}_{p^* p'^* q^*}(\zeta, \zeta', \xi) = \sqrt{4\pi} \int \chi^*_{p^*}(\zeta, \vec{r}_{a1}) \chi_{p'^*}(\zeta', \vec{r}_{c1}) \chi_{q^*}(\xi, \vec{r}_{g1}) dv_1. \tag{13}$$

With the evaluation of integral (13) t is necessary to separate the integration variables from those related to the coordinates of the point $g$. For this purpose we use in Eq. (13) the nonsymmetrical one-range addition theorems for the expansion of multicenter charge densities of STO in the following form [5]:

$$\chi^*_{p_1^*}(\zeta_1, \vec{r}_{a1}) \chi_{p_1'^*}(\zeta_1', \vec{r}_{c1}) = \frac{1}{\sqrt{4\pi}} \lim_{N \to \infty} \sum_{n=1}^{N} \sum_{l=0}^{n-1} \sum_{m=-l}^{l} \begin{bmatrix} W^{\alpha N}_{p_1^* p_1'^* p}(\zeta_1, \zeta_1', z; \vec{R}_{ca}, \vec{R}_{ag}) \\ W^{\alpha N}_{p_1^* p_1'^* p}(\zeta_1, \zeta_1', z; \vec{R}_{cg}, \vec{R}_{ag}) \end{bmatrix} \chi^*_p(z, \vec{r}_{g1}) \tag{14}$$

$$\chi^*_{p_1^*}(\zeta_1, \vec{r}_{a1}) \chi_{p_1'^*}(\zeta_1', \vec{r}_{c1}) = \frac{1}{\sqrt{4\pi}} \lim_{N \to \infty} \sum_{n=1}^{N} \sum_{l=0}^{n-1} \sum_{m=-l}^{l} W^{\alpha N}_{p_1^* p_1'^* p}(\zeta_1, \zeta_1', z; \vec{R}_{ac}, 0) \chi^*_p(z, \vec{r}_{a1}) \tag{15}$$

$$\chi^*_{p_1^*}(\zeta_1, \vec{r}_{a1}) \chi_{p_1'^*}(\zeta_1', \vec{r}_{g1}) = \frac{1}{\sqrt{4\pi}} \lim_{N \to \infty} \sum_{n=1}^{N} \sum_{l=0}^{n-1} \sum_{m=-l}^{l} W^{\alpha N}_{p_1^* p_1'^* p}(\zeta_1, \zeta_1', z; 0, \vec{R}_{ag}) \chi^*_p(z, \vec{r}_{g1}) \tag{16}$$

$$\chi^*_{p_1^*}(\zeta_1, \vec{r}_{b1}) \chi_{p_1'^*}(\zeta_1', \vec{r}_{g1}) = \frac{1}{\sqrt{4\pi}} \lim_{N \to \infty} \sum_{l=|l_1-l_1'|}^{l_1+l_1'} \sum_{m=-l}^{l} W_{p_1^* p_1'^* p^*}(\zeta_1, \zeta_1', z) \chi^*_{p^*}(z, \vec{r}_{g1}), \tag{17}$$

where $p^* \equiv n^*lm$, , $n^* = n_1^* + n_1'^* - 1$, $p \equiv nlm$ and $z = \zeta_1 + \zeta_1'$. See Ref. [9] for the exact definition of multicenter expansion coefficients $W^{\alpha N}$ occurring in Eqs. (14)-(17).

Taking into account Eqs. (14), (15), (16) and (17) in (13) we obtain for one-electron multicenter integrals of NISTO and Coulomb-Yukawa like NICIP the following formulae:

for three-center integrals

$$I^{acg}_{p_1^* p_1'^* q^*}(\zeta_1, \zeta_1', \xi) = \frac{1}{(2\nu+1)^{1/2}} \lim_{N \to \infty} \sum_{n=1}^{N} \sum_{l=0}^{n-1} \sum_{m=-l}^{l} W^{\alpha N}_{p_1^* p_1'^* p}(\zeta_1, \zeta_1' z; \vec{R}_{ca}, 0) \Lambda_{pq^*}(z, \xi, \vec{R}_{ag}) \tag{18}$$

$$I^{acg}_{p_1^* p_1'^* q^*}(\zeta_1, \zeta_1', \xi) = \frac{1}{(2\nu+1)^{1/2}(z+\xi)^{\mu^*+1/2}} \lim_{N \to \infty} \sum_{n=\nu+1}^{N} \begin{bmatrix} W^{\alpha N}_{p_1^* p_1'^*, n\nu\sigma}(\zeta_1, \zeta_1', z; \vec{R}_{ca}, \vec{R}_{ag}) \\ W^{\alpha N}_{p_1^* p_1'^*, n\nu\sigma}(\zeta_1, \zeta_1', z; \vec{R}_{cg}, \vec{R}_{ag}) \end{bmatrix} \frac{\Gamma(n+\mu^*+1)}{[\Gamma(2n+1)]^{1/2}} \left(\frac{2z}{z+\xi}\right)^{n+1/2}, \tag{19}$$

for two-center integrals

$$I^{aag}_{p_1^* p_1'^* q^*}(\zeta_1, \zeta_1'; \xi) = \left(\frac{4\pi}{2\nu+1}\right)^{1/2} \frac{[\Gamma(2\mu^*+1)]^{1/2}}{(2\xi)^{\mu^*+1/2}} \int \chi^*_{p_1^*}(\zeta_1, \vec{r}_{a1}) \chi_{p_1'^*}(\zeta_1', \vec{r}_{a1}) \chi_{q^*}(\xi, \vec{r}_{g1}) dv_1 \tag{20a}$$

$$= \frac{1}{(2\nu+1)^{1/2}} \sum_{l=|l_1-l_1'|}^{l_1+l_1'} \sum_{m=-l}^{l} W_{p_1^* p_1'^* p^*}(\zeta_1, \zeta_1', z) \Lambda_{p^* q^*}(z, \xi; \vec{R}_{ag}) \tag{20b}$$

$$I^{agg}_{p_1^* p_1'^* q^*}\left(\zeta_1,\zeta_1',\xi\right) = \left(\frac{4\pi}{2\nu+1}\right)^{1/2} \frac{\left[\Gamma(2\mu^*+1)\right]^{1/2}}{(2\xi)^{\mu^*+1/2}} \int \chi_{p_1^*}^*\left(\zeta_1,\vec{r}_{a1}\right) \chi_{p_1'^*}\left(\zeta_1',\vec{r}_{g1}\right) \chi_{q^*}\left(\xi,\vec{r}_{g1}\right) dv_1 \quad (21a)$$

$$= \frac{1}{(2\nu+1)^{1/2}(z+\xi)^{\mu^*+1/2}} \lim_{N\to\infty} \sum_{n=\nu+1}^{N} W^{\alpha N}_{p_1^* p_1'^*, n\nu\sigma}\left(\zeta_1,\zeta_1',z;0,\vec{R}_{ag}\right) \frac{\Gamma(n+\mu^*+1)}{\left[\Gamma(2n+1)\right]^{1/2}} \left(\frac{2z}{z+\xi}\right)^{n+1/2} \quad (21b)$$

for one-center integrals

$$I^{ggg}_{p_1^* p_1'^* q^*}\left(\zeta_1,\zeta_1',\xi\right) = C^{\nu|\sigma|}(l_1 m_1, l_1' m_1') A^{\sigma}_{m_1 m_1'} \frac{\Gamma(n_1^* + n_1'^* + \mu^*)}{\left[\Gamma(2n_1^*+1)\Gamma(2n_1'^*+1)\right]^{1/2}} \frac{x_1^{n_1^*+1/2} x_1'^{n_1'^*+1/2}}{y^{\mu^*-1}}, \quad (22)$$

where $z = \zeta_1' + \zeta_1$, $x_1 = \dfrac{2\zeta_1}{y}$, $x_1' = \dfrac{2\zeta_1'}{y}$ and $y = \zeta_1 + \zeta_1' + \xi$. In Eq. (20b), the quantities

$\Lambda_{p^* q^*}(z,\xi,\vec{R}_{ag})$ are the modified overlap integrals of NISTO and Coulomb-Yukawa like NICIP defined as

$$\Lambda_{p^* q^*}(z,\xi;\vec{R}_{ag}) = \frac{\left[\Gamma(2\mu^*+1)\right]^{1/2}}{(2\xi)^{\mu^*+1/2}} S_{p^* q^*}(z,\xi;\vec{R}_{ag}) \quad (23)$$

Here, $S_{p^* q^*}(z,\xi;\vec{R}_{ag})$ are the overlap integrals over NISTO defined by

$$S_{p^* q^*}(z,\xi;\vec{R}_{ag}) = \int \chi_{p^*}^*(z,\vec{r}_a) \chi_{q^*}(\xi,\vec{r}_g) dv \quad (24)$$

As can be seen from Eqs. (8), (11)-(13) and (18)-(24), we have derived a large number of different ($\alpha = 1, 0, -1, -2, ...$) sets of formulae for potential produced by all of the charges of molecule in terms of linear combination coefficients of molecular orbitals and two-center overlap integrals over NISTO. For the calculation of these overlap integrals the efficient computer programs especially useful for large quantum numbers are available in our group [10]. Therefore by using the computer programs for the two-center overlap integrals one can calculate the potential of the field produced by molecule.

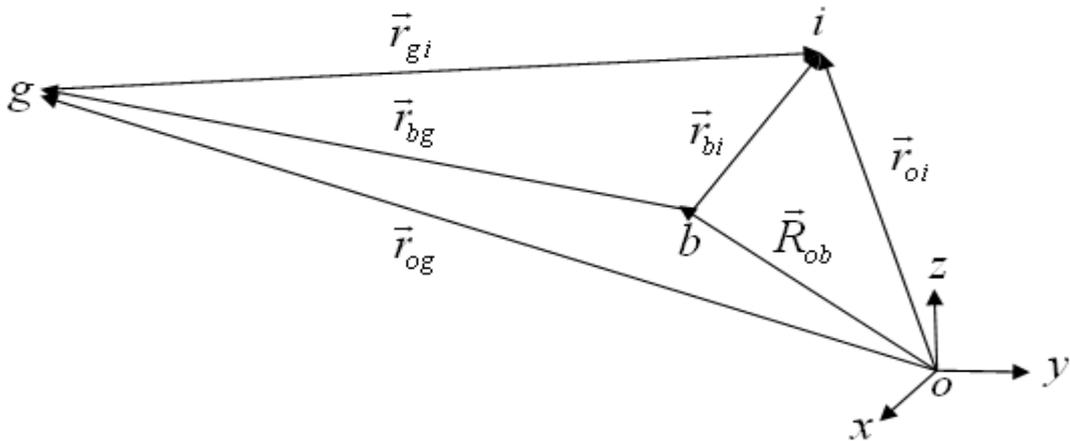

Fig. 1

## 4. Potential energy of interaction between molecules

To find the relation for energy of interaction between molecules we introduce the new lokal coordinate system with its origin $o'$ anywhere within the system of charges of the second molecule (see Fig.2); $\vec{r}_{o'g}$ and $\vec{R}_{o'b'}$ are radius vectors of the electrons and nuclei of this molecule in these coordinates, where $b' \equiv a', c', ...$ In Fig.2, the common coordinate system is denoted by OXYZ. For the determination of the interaction potential energy we regard one of the molecules as being in the field of the other. Then, by the use of Eq. (7), we obtain for the operator of potential energy of the interaction between molecules the relation

$$\hat{U} = \sum_{b'} Z_{b'} \hat{\varphi}(\vec{R}_{ob'}) - \sum_{i'=1}^{N'} \hat{\varphi}(\vec{r}_{oi'}) \tag{25a}$$

$$= \sum_{b'} Z_{b'} \left[ \sum_b Z_b \hat{h}^{q^*}(\xi, \vec{R}_{bb'}) - 2\sum_{i=1}^n f_i \hat{h}_i^{q^*}(\xi, \vec{R}_{ob'}) \right] - \sum_{i'=1}^{N'} \left[ \sum_b Z_b \hat{h}^{q^*}(\xi, \vec{r}_{bi'}) - 2\sum_{i=1}^n f_i \hat{h}_i^{q^*}(\xi, \vec{r}_{oi'}) \right]. \tag{25b}$$

The everage expectation value of operator $\hat{U}$ should be calculated with the help of orthonormal sets of multideterminantal wave functions of second molecule. Carrying through calculations for the second molecule analagous to those for the first one, we obtain for the potential energy of interaction between molecules the following relation:

$$U = U_1 + U_2 + U_3 + U_4, \tag{26}$$

where $U_1$, $U_2$, $U_3$ and $U_4$ are the average values of potential energies of interactşon between partricles of molecules:
for nuclei-nuclei

$$U_1 = \sum_{b'} \sum_b Z_{b'} Z_b h^{q^*}(\xi, \vec{R}_{bb'}), \tag{27}$$

for electrons-nuclei

$$U_2 = -2 \sum_{b'} \sum_{i=1}^n Z_{b'} f_i h_i^{q^*}(\xi, \vec{R}_{ob'}) \tag{28}$$

$$U_3 = -2 \sum_{i'=1}^{n'} \sum_b Z_b f_{i'} h_{i'}^{q^*}, \tag{29}$$

for electrons-electrons

$$U_4 = 4 \sum_{i'=1}^{n'} \sum_{i=1}^n f_{i'} f_i h_{i'i}^{q^*}. \tag{30}$$

Here, the quantities $h^{q^*}$ and $h_i^{q^*}$ are determined by Eqs (2) and (9), respectively, and

$$h_{i'}^{q^*} = \int u_{i'}^*(\vec{r}_2) \hat{h}^{q^*}(\xi, \vec{r}_{b2}) u_{i'}(\vec{r}_2) dv_2 \tag{31}$$

$$h_{i'i}^{q} = \int u_{i'}^*(\vec{r}_2) \hat{h}_i^{q^*}(\xi, \vec{r}_{o2}) u_{i'}(\vec{r}_2) dv_2 \tag{32a}$$

$$= \int u_{i'}^*(\vec{r}_2) \left( \int u_i^*(\vec{r}_1) \hat{h}^{q^*}(\xi, \vec{r}_{21}) u_i(\vec{r}_1) dv_1 \right) u_{i'}(\vec{r}_2) dv_2, \tag{32b}$$

where $u_{i'}$ are the molecular orbitals of the second molecule defined as

$$u_{i'} = \sum_{p_2^*} \chi_{p_2^*} C'_{p_2^* i'}. \tag{33}$$

Substituting (33) into Eqs. (31) and (32) we obtain:

$$h_{i'}^{q^*} = \sum_{p_2^* p_2'^*} C'^*_{p_2^* i'} C'_{p_2'^* i'} \Lambda_{p_2^* p_2'^* q^*}^{a'c'b'}(\zeta, \zeta', \xi) \tag{34}$$

$$h^{q^*}_{i'i} = \sum_{p_2^* p_2'^*} C'^*_{p_2^* i'} C'_{p_2'^* i'} \sum_{p_1^* p_1'^*} C^*_{p_1^* i} C_{p_1'^* i} I_{p_2^* p_2'^*, p_1^* p_1'^*, q^*}, \tag{35}$$

where

$$I_{p_2^* p_2'^*, p_1^* p_1'^*, q^*} = \int \chi^*_{p_2^*}(\zeta_2, \vec{r}_{a'2}) \left( \int \chi^*_{p_1^*}(\zeta_1, \vec{r}_{a1}) \chi_{p_1'^*}(\zeta_1', \vec{r}_{c1}) h^{q^*}(\xi, \vec{r}_{21}) dv_1 \right) \chi_{p_2'^*}(\zeta_2', \vec{r}_{c'2}) dv_2. \tag{36}$$

Now we take into account the symmetrical one-range addition theorems for the Coulomb-Yukawa like NICIP (see Eq. (20) of Ref. [6]). Then, the integral (36) can be expressed through the products of the following integrals:

for first molecule

$$I^{acO}_{p_1^* p_1'^* q_1}(\zeta_1, \zeta_1', \xi) = \int \chi^*_{p_1^*}(\zeta_1, \vec{r}_{a1}) \chi_{p_1'^*}(\zeta_1', \vec{r}_{c1}) \chi_{q_1}(\xi, \vec{r}_{O1}) dv_1, \tag{37}$$

for second molecule

$$I^{a'c'O}_{p_2^* p_2'^* q_2}(\zeta_2, \zeta_2', \xi) = \int \chi^*_{p_2^*}(\zeta_2, \vec{r}_{a'2}) \chi_{p_2'^*}(\zeta_2', \vec{r}_{c'2}) \chi_{q_2}(\xi, \vec{r}_{O2}) dv_2, \tag{38}$$

where $q_i \equiv \mu_i \nu_i \sigma_i$; $\vec{r}_{O1}$ and $\vec{r}_{O2}$ are the radius vectors with respect to the common coordinate system OXYZ. The multicenter overlap integrals (37) and (38) can be calculated by the use of formulae (18)-(24).

As can be seen from equations of this section, the two-center overlap integrals and the linear combination coefficients of molecular orbitals occur in the potential energy of interaction. Using the values of these quantities one can calculate the interaction potential energy between arbitrary atomic-molecular systems which have any number of closed and open shells.

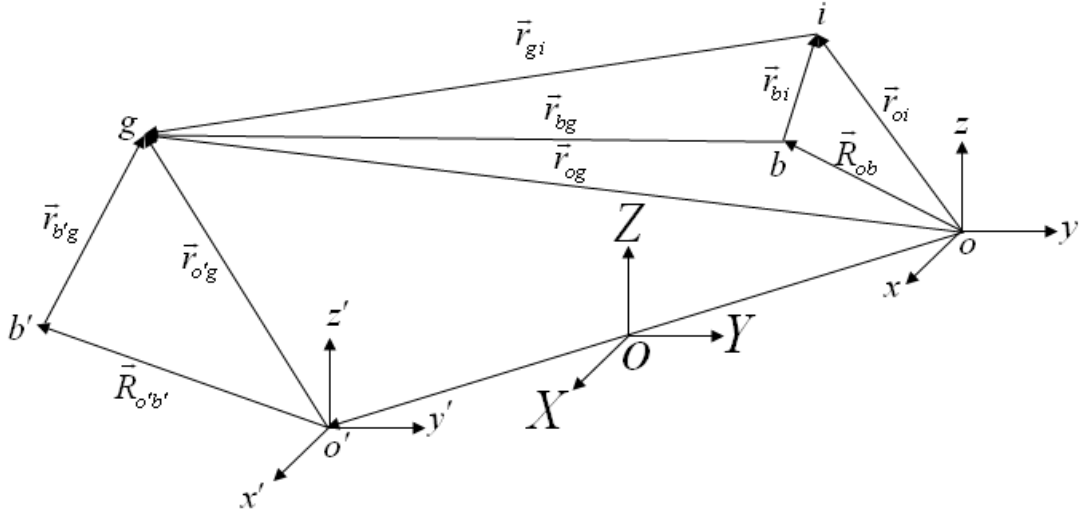

**Fig.2**